# Mechanical and Thermal Stability of Graphyne and Graphdiyne Nanoscrolls

Daniel Solis[1], Cristiano F. Woellner[1,2], Daiane D. Borges[1], and Douglas S. Galvao[1]
[1]Applied Physics Department, University of Campinas - UNICAMP, Campinas-SP 13083-959, Campinas-SP, Brazil
[2]Department of Materials Science and Nano Engineering, Rice University, Houston, Texas, USA

## ABSTRACT

Graphynes and graphdiynes are carbon 2D allotrope structures presenting both sp$^2$ and sp hybridized atoms. These materials have been theoretically predicted but due to intrinsic difficulties in their synthesis, only recently some of these structures have been experimentally realized. Graphyne nanoscrolls are structures obtained by rolling up graphyne sheets into papyrus-like structures. In this work, we have investigated, through fully atomistic reactive molecular dynamics simulations, the dynamics of nanoscroll formation for a series of graphyne (α, β, and δ types) structures. We have also investigated their thermal stability for a temperature range of 200-1000K. Our results show that stable nanoscrolls can be formed for all structures considered here. Their stability depends on a critical value of the ratio between length and height of the graphyne sheets. Our findings also show that these structures are structurally less stable then graphene-based nanoscrolls. This can be explained by the graphyne higher structural porosity which results in a decreased pi-pi stacking interactions.

## INTRODUCTION

With the materials science revolution created by the advent of graphene, there is a renewed interest in other 2D carbon allotropes, such as graphynes and graphdiynes [1-5] (Figure 1). Graphynes and graphdiynes are carbon 2D allotrope forms presenting both sp2 and sp carbon hybridized atoms and many different structures can be generated form these motifs [1-5]. The main difference between graphynes and graphdiynes are the number of acetylenic (one and two, respectively) groups connecting benzene rings. These materials have been theoretically predicted in 1987 [1], but due to intrinsic difficulties in their synthesis, only recently some structures have been experimentally realized [6,7]. These successful syntheses renewed the interest in these structures and remarkable electronic properties have been predicted for some of them [1-8]. Graphyne-like nanotubes [2] have already been successfully produced [7], but other structures remain to be explored. Among these potential new structures is worth mention the possibility of the existence of graphyne/graphdiyne nanoscrolls. Nanoscrolls are structures obtained by rolling up nanosheeets into papyrus-like topology [9] (Figure 1). Nanoscrolls present high radial flexibility and large solvent accessible surface area, which opens the possibility of many applications. This kind of structure has been already experimentally realized with different materials, such as graphene, graphene oxide and hexagonal boron nitride, for a recent scroll review see ref. [10].

In this work, we have investigated, through fully atomistic reactive molecular dynamics simulations, the dynamics of nanoscroll formation for a series of α, β, and δ graphyne and graphdiyne nanoscrolls. We have investigated their dynamics formation and their structural and thermal stability for a temperature range of 200-1000K.

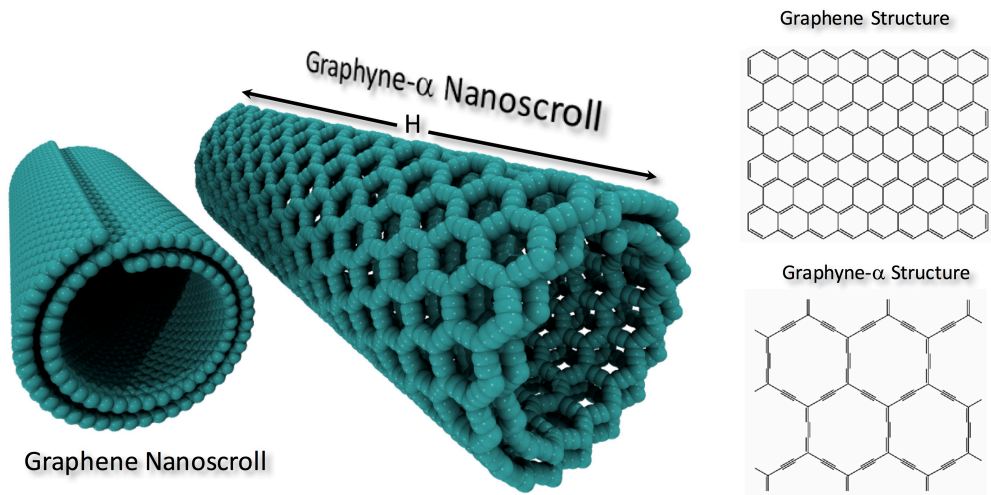

**Figure 1**. Graphene and α-graphyne scrolls formed from rolling up graphene and α-graphyne sheets, respectively. From a topological point of view α-graphyne can be considered as a graphene structure where acetylenic (triple bonds) groups were inserted into each hexagonal bond, which results in a significant increase in structural porosity. H indicates the nanoscrolls width (graphene/graphyne/graphdiyne).

**THEORY**

We have carried out fully atomistic molecular dynamics (MD) simulations using the reactive (allows breaking and chemical bond formation) force field REAXFF [11], as implemented in the open source code large-scale atomic/molecular massively parallel simulator (LAMMPS) [12]. We used a NVT ensemble, with the target temperature controlled by a Nosé−Hoover thermostat [12]. The temperature range investigated here was from 0 up to 1000 K. The time step was chosen to be 0.2 fs (femtoseconds) to ensure the numerical stability of the simulations.
   (i) Static simulations, which was used to analyze the energetic cost of the scroll formation from rolling up the nanosheets with typical dimensions of 100 Å (H) x 300 Å (W). H and W indicate the nanoscroll height (see Figure 1) and length, respectively. See the inset "Initial configuration" in Figure 2.
- (ii) Dynamics simulations. In this part, we performed fully atomistic molecular dynamics simulations to investigate the thermal scroll stability at different temperatures.

**DISCUSSION**

In Figure 2 we present the variation of potential energy for α-graphyne as function of the inner radius for three different sheet lengths: W=100, 200 and 300 Å, while keeping the same height, H=100 Å, in order to investigate how the scroll stability depends on its length dimensions. The potential energy for the planar configuration (see "initial configuration" in Figure 2) was taken as the reference energy value (equals to zero). The geometrical meaning of H (height) and W (length) can be seen in Figure 1 and 2, respectively.

The scroll formation is a competition between the cost of elastic energy to bend the sheet (which needs to be energy assisted) and the structural gain resulting from the van der Waals energy of the scrolling process. If this balance is negative the scrolled structure would be stable and it is characterized by an energy minimum below zero in Figure 2. As can see in Figure 2, for the case of W=100 Å, the nanoscroll is energetically unfavorable as its potential minimum value energy is larger than the corresponding planar configuration. Comparing bending and van der Waals energy interactions we can observe that in this case the bending energy has greater contributions than van der Walls. For the W values of 200 and 300 Å, stable scroll configurations were obtained.

Increasing the sheet length (W) results in stronger van der Waals interactions due to the increased overlapping surface area, which favors the scroll formation. As expected the scroll stability depends on its dimensions (W and H values). For space limitation, we restrict ourselves only to analyze the stability on the length (W) dimension, the height (H) dependence was investigated for other scroll materials [10] and the same conclusions hold here.

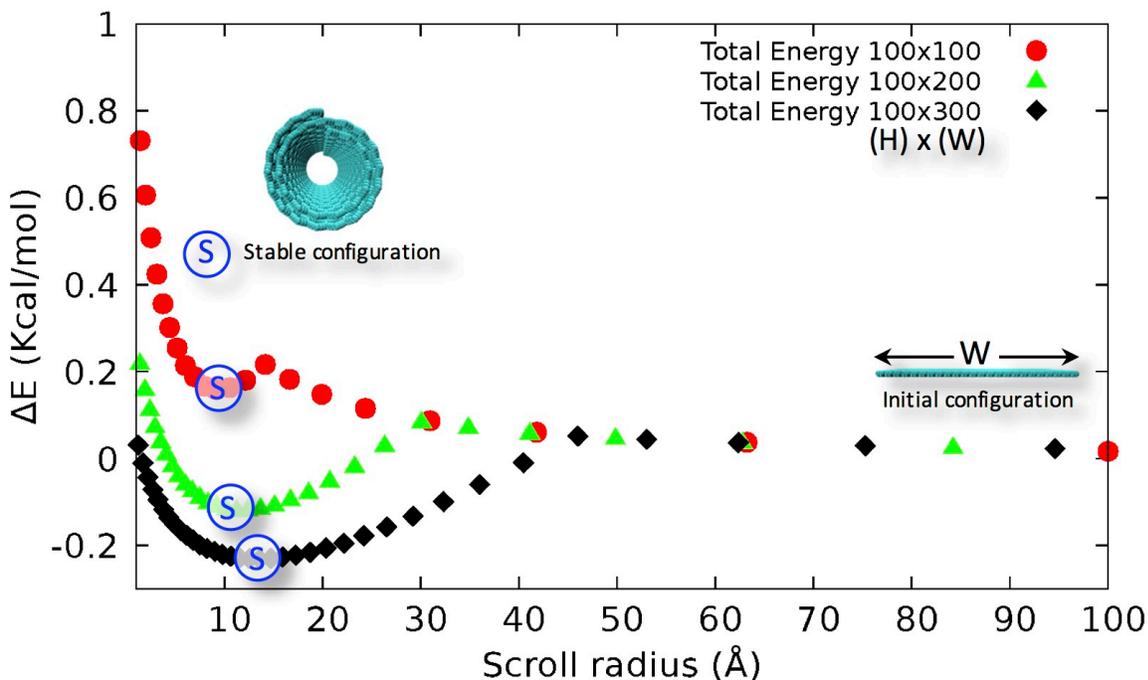

**Figure 2**. Potential energy variation as a function of the inner radius for three different nanoscroll lengths (W). The nanoscrolls height is kept constant and equals to H=100 Å. The geometrical mean of nanoscroll length (W) is depicted in "initial configuration" inset. And for the nanoscroll height (H) see Figure 1. The "S" symbol refers to the position of the minimum energy value and its inner radius value. Notice that for the square sheet (solid circle symbols) the nanoscroll configuration is energetically unfavorable to occur.

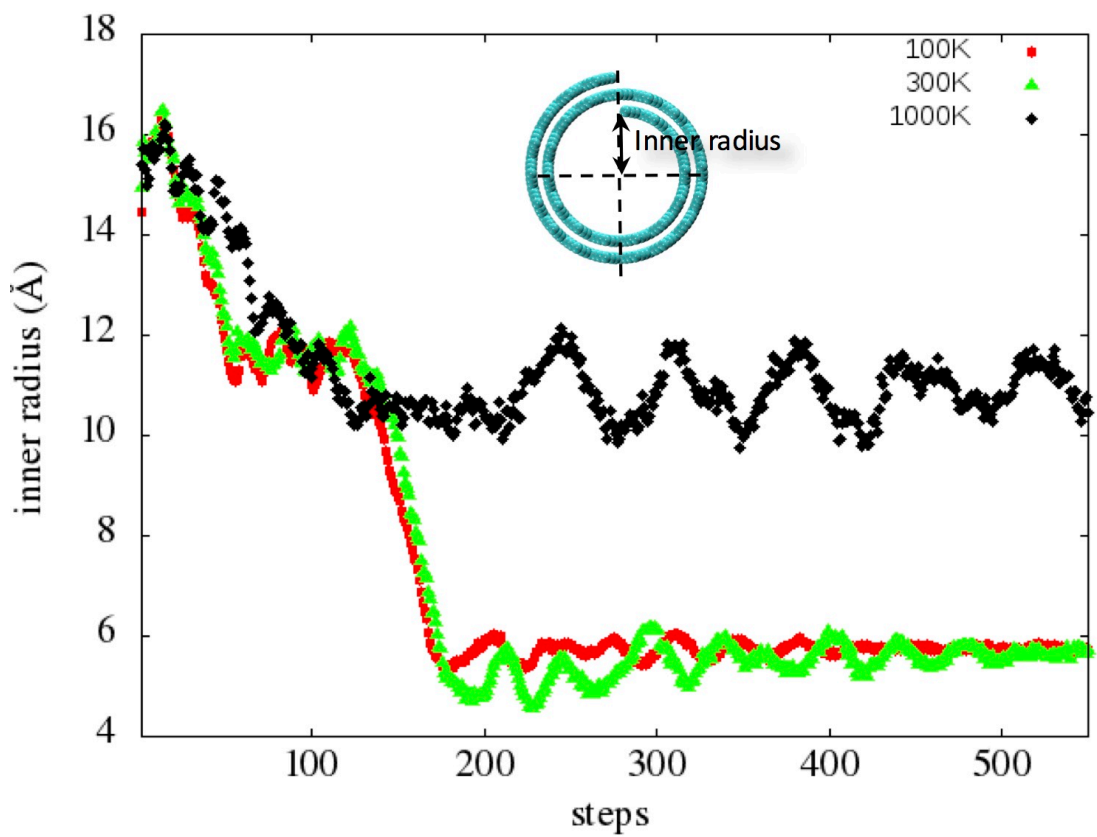

**Figure 3.** Typical time evolution of the scroll inner radius values for temperatures of 100, 300 and 1000K, respectively. Notice the strong inner radius dependence on the temperature values.

In Figure 3 we present how the scroll inner radius values varies as a function of simulation time for three different temperatures. The MD simulations were carried out starting from an initial configuration of 4 pi (rounds), inner radius around 15Å and inner-to-outer distance around 7Å, the structures were then set free to evolve in time. As we can see from this figure, although the scroll inner radius values oscillate in time (scroll 'breathings'), for all temperatures investigated here, thermally stable configurations were obtained. However, the thermal asymptotic behavior for the inner radius value has a strong dependence on the temperature values, which can be exploit as the basis for building thermal actuators [13].

This can be better evidenced in Figure 4, where we present MD snapshots of thermalized α-graphyne nanoscrolls at three different temperatures. Two important results can be inferred from this figure: α-graphyne nanoscrolls are stable for a large range of temperatures (at least up to 1000K) and the thermalized inner radius values depends on the temperature. As mentioned above this characteristic open a new horizon for applications in graphyne nanoscrolls based materials. For instance, tuning these values could be the basis of thermal actuators [13].

Figure 5 shows MD snapshots of α, β and γ– graphyne and graphdiyne nanoscrolls. These snapshots represent the thermalized (stable) structures at 100 K after 200 ps. Clearly from this figure, we can see that for each series: $k=α, β$ and γ; $k$-graphdiyne has larger inner radius as compared to $k$-graphyne. It can be attributed to their difference in porosity as follow: as mentioned earlier, the main structural difference between graphynes and graphdiynes are the number of acetylenic groups present in the structures. Graphdiyne presents twice as much acetylenic groups as compared to graphyne. Therefore, graphdiyne is a more porous structure then graphyne. More porosity results in a decreased pi-pi stacking interactions, as discussed above.

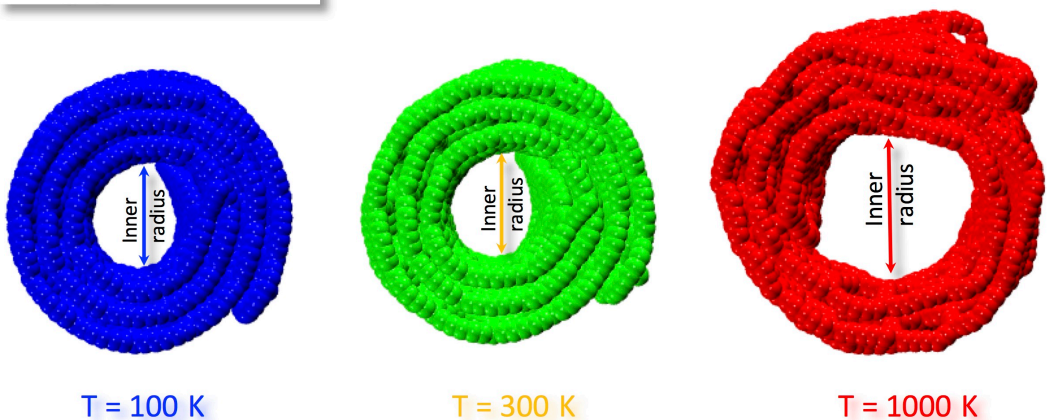

**Figure 4.** MD snapshots of nanoscrolls at 100, 300, and 1000K, respectively. These snapshots represent thermalized structures at each respective temperature. Results for α-graphyne.

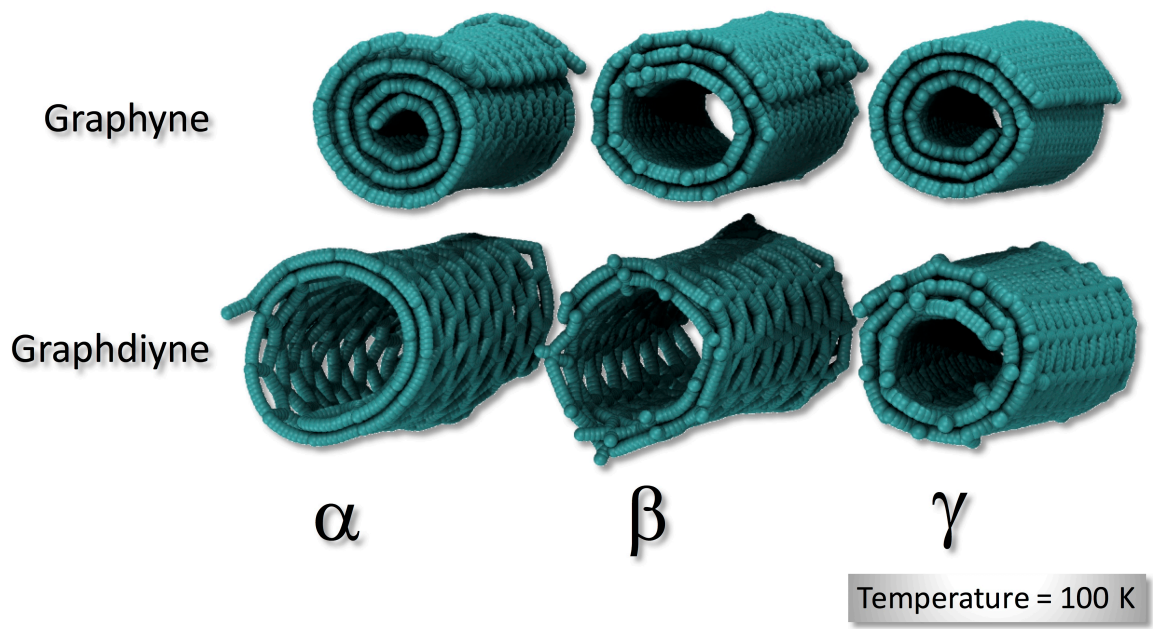

**Figure 5.** Snapshots from MD simulations of α, β, γ graphyne (upper panel) and graphdiynes (lower panel) nanoscrolls at 100 K, respectively. These snapshots represent thermalized structures after 200 ps.

## CONCLUSIONS

In this work using fully atomistic reactive molecular dynamics simulations we investigated the dynamics and thermal stability (up to 1000 K) of graphyne and graphdiyne nanoscrolls. Our results show that thermally stable (at least up to 1000K) nanoscrolls can be formed for all structures considered here. Their stability depends on a critical value of the ratio between length and height of the graphyne sheets as showed in Figure 2. Our findings also show that these structures are structurally less stable then graphene-based nanoscrolls. This can be explained by the graphyne higher structural porosity which results in a decreased pi-pi stacking interactions. We have also shown, for α-graphyne case, that the scroll internal radius values can be thermal tuned, which can be exploited to use these nanostructures as thermal actuators. Due to space limitations, only the results for α-graphyne the results were discussed in details, but the general trends and main conclusions are also valid to α, β, and δ graphyne and graphdiyne nanoscrolls (Figure 5).

## ACKNOWLEDGMENTS

This work was supported in part by the Brazilian Agencies CAPES, CNPq and FAPESP. The authors also thank the Center for Computational Engineering and Sciences at Unicamp for financial support through the FAPESP/CEPID Grant # 2013/08293-7. CFW thanks São Paulo Research Foundation (FAPESP) Grant # 2016/12340-9 for financial support.

## REFERENCES

1. R. H. Baughman, H. Eckhardt and M. Kertesz, *J. Chem. Phys.* **87**, 6687 (1987).
2. V. R. Coluci, S. F. Braga, S. B. Legoas, D. S. Galvao, and R. H. Baughman, *Phys. Rev. B* **68**, 035430 (2003).
3. D. Malko, C. Neiss, F. Vines, and A. Gorling, *Phys. Rev. Lett.* **108**, 086804 (2012).
4. P. A. S. Autreto, J. M. de Sousa, and D. S. Galvao, *Carbon* **77**, 829 (2014).
5. J. M. de Sousa, G. Brunetto, V. R. Coluci and D. S. Galvao, *Carbon* **96**, 14 (2016).
6. G. Li, Y. Li, H. Guo, Y. Li and D. Zhu, *Chem. Commun.* **46**, 3256 (2010
7. G. Li, Y. Li, X. Qian, H. Liu, H. Lin, N. Chen, and Y. Li, *J. Phys. Chem. C* **115**, 2611 (2011).
8. A. N. Enyashin and A. L. Ivanovskii, *Phys. St. Solid B,* 1 (2011).
9. S. F. Braga, V. R. Coluci S. B. Legoas, R. Giro, D. S. Galvao, and R. H. Baughman, *Nano Lett.* **4**, 881 (2004).
10. E. Perim, L. D. Machado and D. S. Galvao, Frontiers in Materials, **1**, 31 (2014).
11. A. C. T. Van Duin, S. Dasgupta, F. Lorant, W. A. Goddard, J. Phys. Chem. A, 105, 9396 (2001)
12. S. J. Plimpton, Comput. Phys., 117, 1 (1995). http://lammps.sandia.gov (accessed on 01/06/2017).
13. J. Loomis, X. Fan, F. Khosravi, P. Xu, M. Fletcher, R. W. Cohn, and B. Panchapakesan, *Sci. Rep.* **3**, 1900 (2013).